\documentclass[amsmath,superscriptaddress,showpacs,prb,twocolumn] {revtex4}
\usepackage{bm}
\usepackage{enumerate}
\usepackage{graphicx}
\usepackage[dvips]{epsfig}

\makeatletter
\newcommand{\Rmnum}[1]{\expandafter\@slowromancap\romannumeral #1@}
\makeatletter

\begin{document}
\title{Role of domain wall fluctuations in non-Fermi liquid behavior of metamagnets}
\author{Vladimir A. Zyuzin}
\affiliation{Department of Physics, The University of Texas at Austin, Austin, TX 78712, USA}
\author{A.Yu. Zyuzin}
\affiliation{A.F. Ioffe Physical-Technical Institute, 194021, St.Petersburg, Russia}
\begin{abstract}
In this paper we study resistivity temperature dependence of a three dimensional metamagnet near the metamagnet phase transition point. The phase transition is characterized by a phase separation of regions with high and low magnetization. We show that in the case of weak pinning, the spin relaxation time
of the domain wall, which separates the two phases, is much larger than that of the volume spin fluctuations. This opens a temperature range where resistivity temperature dependence is determined by
scattering of conducting electrons by the domain wall fluctuations. We show that it leads to quasi-linear low temperature dependence of resistivity.

\end{abstract}
\maketitle

\section{Introduction}

Understanding the deviation from the Fermi liquid behavior at phase transition critical points is a current research question. The most important are deviations from quadratic temperature dependence of resistivity near quantum critical points. Theory proposes that the origin of non-Fermi liquid behavior is scattering of conducting electrons by bosonic critical soft modes.\cite{Moriya, Hertz, Millis} In case of ferromagnetic transition these modes are collective spin fluctuations whose relaxation time diverges near the transition point. In nearly magnetic metals, at temperatures larger than the inverse spin relaxation time, contribution to the resistivity due to electron scattering by spin fluctuations strongly deviates from Fermi liquid quadratic dependence. It becomes linear or even saturates.\cite{Shindler, Ueda, Stewart} In strongly Stoner enhanced paramagnetic metals the deviation might start at very low temperature.

It was experimentally found that in the region near putative quantum critical point, the magnetic state of many systems experiences broadened first order phase transition with separation of different phases.\cite{Goto, Nagaev, Lonzarich, Zamborszky, Pfleiderer, Grigera, Perry,  Uhlarz, Mao, Uemura, ZrZn2}. In these experiments, the temperature dependence of resistivity is characterized by non-Fermi liquid exponents, which depend on proximity of the system to the putative quantum critical point. Since at the first order phase transition the spin fluctuations are not critical, the theory of scattering of conducting electrons by spin fluctuations \cite{Moriya, Hertz, Millis}, discussed above, can not be directly applied to describe the non-Fermi liquid contributions to the resistivity. Alternative theories were developed to overcome some difficulties \cite{Belitz_prl99, Millis_prl02}, but complete understanding is lacking. \cite{Coleman}

The phase separation, which is a natural phenomena occurring at the first order phase transition, suggests an importance of considering the contributions to the resistivity due to structural differences of separated phases and phase boundaries (domain walls).\cite{Dikshtein_jetp84, Dikshtein_jetp86} We would like to point out that the nonuniform magnetic state can be considered as a distinctive phase itself. \cite{Nagaev, Dikshtein_jetp84, Dikshtein_jetp86, Binz, Berridge_prl09, Maleev, Conduit} For example, at the metamagnetic transition of thin magnetic films, the magneto-dipole interaction can give rise to formation of magnetic domains of various types.\cite{Dikshtein_jetp84, Dikshtein_jetp86, Binz} Also, a magnetic analog of a Fulde-Ferrel-Larkin-Ovchinnikov phase can develop at the first order phase transition.\cite{Berridge_prl09} In another example, the domains can be formed in spatially random magnetic field. This model was proposed to explain unusual magnetotransport effects of ferromagnetic alloys.\cite{Burkov}

In this paper we propose a possible scenario of appearance of bosonic critical soft modes at the magnetic first order phase transition driven by a magnetic field. The system of our study is a three-dimensional metamagnet that splits into regions with high and low magnetization (magnetic domains) with corresponding domain walls. We show that the spin fluctuations, which are short lived inside the magnetic domains, might become critical at the domain walls. Therefore, in the spirit of theory \cite{Moriya, Hertz, Millis}, the scattering of conducting electron by these domain wall fluctuations results in non-Fermi liquid behavior of resistivity.

The rest of the paper is organized as following. In section \Rmnum{2} we give a description of the three-dimensional system of coupled conducting electrons and itinerant electrons responsible for metamagnetic state. We propose a model where system splits in to domain walls due to spacial variation of magnetic field. In section \Rmnum{3} the solution for the domain wall is obtained. The dynamical magnetic susceptibility of the domain wall fluctuations is then derived. In section \Rmnum{4} the temperature dependence of the resistivity caused by the scattering of conducting electrons by the domain wall fluctuations is considered. We find transition from quadratic to linear temperature dependence with increasing temperature. We give a condition when this contribution is larger than that from the volume spin fluctuations. \cite{Shindler, Ueda, Moriya} In section \Rmnum{5} we discuss the so-called interaction correction\cite{AlAr} to the conductivity at small temperatures. It is known that scattering of conducting electrons by impurities and spin fluctuations results in important contributions to the temperature dependence of conductivity.\cite{Rosch, Kim_Millis, Paul, Kirkpatrick} We find that for the domain wall fluctuations, the interaction correction to the conductivity has a temperature dependence predicted for the two-dimensional electron system close to ferromagnetic quantum critical point.\cite{Paul}

\section{Description of the model}
We consider a three dimensional metallic metamagnet where $s-$ electrons are considered to be conducting, and $d-$ electrons are responsible for the magnetic state. Coupling of conducting electrons with bosonic modes is obtained by integrating out $d-$electrons.\cite{Abanov}
Action describing conducting electrons in random impurity potential $V({\bf r})$ is
\begin{equation}\label{action_free}\small
S_{s}= T\sum_{\Omega_{n}}\int d{\bf r}~ \psi_{\alpha}^{\dag}({\bf r}, \Omega_{n})\left[i\Omega_{n} + \frac{\nabla^{2}}{2m_{e}} +\mu - V({\bf r}) \right]\psi_{\alpha}({\bf r},\Omega_{n}),
\end{equation}
where $\psi_{\alpha}$ describes electrons with mass $m_{e}$, spin $\alpha$, Matsubara frequency $\Omega_{n}=\pi T(2n+1)$ ($T$ is temperature), with Fermi level $\mu$, and we have set $\hbar=1$. We assume impurity potential to satisfy $<V({\bf r})>=0$, and $<V({\bf r})V({\bf r}^{\prime})>=1/(2\pi\nu\tau)\delta({\bf r}-{\bf r}^{\prime})$, where $\nu$ - density of electrons per one spin, $\tau$ - is the electron mean free time.

The magnetic part of free energy has two energetically unequal minima. Under application of magnetic field  the system undergoes metamagnetic phase transition when these two minima have the same energy. The value of the magnetic field at which it occurs is called the metamagnetic magnetic field, let us denote it by $h_{m}$. Near the metamagnetic phase transition we approximate the free energy by two parabolas with minima at $m(\textbf{r},\tau)=\pm m_{0}$, corresponding to high and low magnetization states. The domain walls originate from the spatial deviation of the magnetic field from its metamagnetic value. We denote such deviation as $h \left({\bf r}\right)$. Assuming a strong magnetic field, we consider only longitudinal component of magnetization density $m({\bf r})$ (in units of $g\mu_{B}\equiv 1$) in the action. Under such assumptions the action describing the magnetization density $m(\textbf{r},\tau)$ has the form
\begin{eqnarray}\label{action}
&S_{mag}[m]=\int_{0}^{1/T}d\tau\int d\mathbf{r}\left(\frac{K}{2\chi_{0}}\left(\mathbf{\nabla}m\left(\mathbf{r},\tau\right)\right)^{2} \right. \nonumber\\ &\left.+ \frac{\alpha}{2\chi_{0}}
(\lvert m\left(\mathbf{r},\tau\right)\rvert-m_{0})^{2}-  h
\left(\mathbf{r}\right) m\left(\mathbf{r},\tau\right)\right)+S_{D},
\end{eqnarray}
here coefficient $K^{-1/2}$ is of the order of electrons Fermi wave length, $\alpha^{-1}$ is Stoner enhancement factor, and $\chi_{0}$ is noninteracting electron spin susceptibility, which we put to be the same for both high and low magnetization phases. This assumption greatly simplifies further calculations, and does not affect main conclusions about resistivity temperature dependence. In our model, the spatial distribution of the magnetic field $h({\bf r})$ has a Gaussian form (see appendix for definition and an averaging procedure of appropriate quantities).

The last term in the action (\ref{action}) is the Landau damping, which is described by
\begin{equation}\label{damping}
S_{D}[m]=\int d{\bf r}^{\prime}d{\bf r}~ T\sum_{\omega_{n}} m\left({\bf r},\omega_{n}\right)\Gamma({\bf r}-{\bf r}^{\prime},|\omega_{n}|) m\left({\bf r}^{\prime},\omega_{n}\right),
\end{equation}
here $\omega_{n}$ is the bosonic Matsubara frequency. At this point it is necessary to distinguish two cases of the damping depending on the regimes of electron scattering by random impurity potential $V({\bf r})$ given in the expression (\ref{action_free}). For ballistic electrons, the Fourier image of $\Gamma({\bf r},|\omega_{n}|)$ is given by $\Gamma({\bf Q},|\omega_{n}|)=\frac{\gamma|\omega_{n}|}{v_{F}Q}$, which is valid for large momenta $v_{F}Q>|\omega_{n}|$ where $v_{F}$ is Fermi velocity, and $\gamma$ is a damping constant. \cite{Moriya, Jull} In case of diffusive electrons, the damping is $\Gamma({\bf Q},|\omega_{n}|)=\frac{\gamma|\omega_{n}|}{DQ^2}$ which is valid for $DQ^2>|\omega_{n}|$ where $D$ is the diffusion constant. \cite{Fulde}

Finally, part of action that describes the coupling of conducting electrons with magnetization is
\begin{equation}\label{coupling}
S_{int}=G\int_{0}^{1/T}d\tau\int d\mathbf{r}~ s\left(\mathbf{r},\tau\right)
m\left(\mathbf{r},\tau\right),
\end{equation}
here ${\bf s}\left(\mathbf{r},\tau\right)$ is operator of spin density of conducting electrons along the longitudinal component of magnetization, $G$ is a phenomenological coupling constant.

\section{Domain wall fluctuations}
In this section we present a solution of the mean field equation for the domain wall and discuss fluctuations around it. Let $x$ be a coordinate normal to the domain wall and $h\left(x_{0} ({\bm\rho})\right)=0$, so that at $x<x_{0}$ it is state with high magnetization, and at $x>x_{0}$ it is a low magnetization state. Here ${\bm \rho}$ is a two-dimensional coordinate along the domain wall. Varying the action (\ref{action}), we obtain equation for the magnetization
\begin{equation}\label{mean-field}
-K\frac{d^{2}} {dx^{2}}m(x)+\alpha \left( m(x)-m_{0}sign (m(x))\right)
=\chi_{0}h(x,{\bm\rho}).
\end{equation}

When $h({\bf r})$ is slowly varying function on a scale of $\sqrt{K/\alpha}$ the domain wall can be approximated as flat.
With this assumption the solution of (\ref{mean-field}) describing the domain wall is
\begin{equation}
m=-m_{0}(1- e^{ -\sqrt{\alpha/K}| x-x_{0}({\bm \rho})|} )sign
(x-x_{0}({\bm \rho})) + \frac{\chi_{0}h(x,{\bm\rho})}{\alpha}.
\end{equation}

Let us consider fluctuations near this solution. Taking second derivative of free energy we obtain equation for eigenfunctions, which describe fluctuations
\begin{eqnarray}\label{delta}
&(-K\nabla ^{2} +\alpha) \delta m({\bf r})& \nonumber\\
&-\frac{2\alpha m_{0}}{\left\vert \frac{d}{dx}(m(x))\right\vert
}_{x=x_{0}}\delta (x-x_{0}({\bm \rho}))\delta m({\bf r})=\epsilon \delta m({\bf r}).
\end{eqnarray}
Deriving this equation we have used the equality $\delta(m(x))=\delta (x-x_{0}({\bm\rho}) )/\left\vert \frac{d}{dx}(m(x))\right\vert_{x=x_{0}}$. The delta-type potential in the equation (\ref{delta}) is related to non analytical dependence of free energy on magnetization. Equation (\ref{delta}) has only one bounded solution, thus strongly simplifying consideration of fluctuations. With an assumption of slowly varying $x_{0}({\bm\rho})$ and $\frac{d}{dx}m(x_{0}({\bm\rho}))$, one can search for the solution of equation (\ref{delta}) in the form of a plane wave in ${\bm\rho}$
\begin{equation}
\delta m(r)=\Psi_{0}(x-x_{0}({\bm \rho}) )e^{i{\bf Q}{\bm \rho}},
\end{equation}
where
\begin{equation}\label{psi-0}
\Psi_{0}(x) =\sqrt{\beta}e^{ -\beta\left\vert x-x_{0}\right\vert },
\end{equation}
and with $\epsilon =KQ^{2}+\epsilon_{0}$, where ${\bf Q}$ is now a two-dimensional wave vector along the domain wall, we write
\begin{equation}
\epsilon_{0} =\alpha \left[ 1-\left( 1+\frac{\chi_{0}}{
m_{0}\sqrt{K\alpha}} \left\vert \frac{d}{dx}h(x) \right\vert
\right) ^{-2}\right],
\end{equation}
where $\beta=\sqrt{\left( \alpha -\epsilon_{0}\right) /K}$. At slowly varying $h(\textbf{r})$, the eigenvalue is small $\epsilon_{0} << \alpha$ and equal
\begin{equation}\label{lifetime}
\epsilon_{0}=\frac{2\chi_{0}}{m_{0}\beta}\left\vert \frac{d}{dx}h(x) \right\vert.
\end{equation}

Dynamics of spin fluctuations is governed by the Landau damping (\ref{damping}) which for excitation described by $\Psi_{0}(x)$ translates to
\begin{equation}\small
\Gamma(Q,|\omega_{n}|)=\frac{\gamma|\omega_{n}|}{v_{F}}\int_{-\infty}^{\infty}dxdx'
\int_{-\infty}^{\infty}dq\frac{e^{i q(x-x')}\Psi_{0}(x)\Psi_{0}(x')}{\sqrt{Q^2+q^2}},
\end{equation}
where as an example we used ballistic case. At small momenta $\beta >Q$ we have for the ballistic case
\begin{equation}\label{ballistic damping}
\Gamma(Q,|\omega_{n}|)=\frac{4\gamma|\omega_{n}|}{\pi \beta v_{F}}\ln(\beta/Q).
\end{equation}
Same procedure for the diffusive case gives
\begin{equation}\label{diffusive damping}
\Gamma(Q,|\omega_{n}|)=\frac{2\gamma|\omega_{n}|}{DQ\beta}.
\end{equation}
The dynamical susceptibility of one domain wall fluctuations is represented in the form
\begin{equation}\label{dyn-sus}
\chi\left(  {\bf r},{\bf r}^{\prime},\omega \right) =\int\frac {d^{2}Q}{\left(2\pi\right)^{2}} e^{
-i\mathbf{Q}\left( {\bm \rho}-{\bm \rho}^{\prime}\right)
}  \Psi_{0}\left( x\right)  \Psi_{0}\left( x^{\prime}\right)
\chi({\bf Q},\omega),
\end{equation}
where
\begin{equation}\label{chi-omega}
\chi({\bf Q},\omega) = \frac{\chi_{0}}{\epsilon_{0}+KQ^{2}+\Gamma(Q,i\omega) }.
\end{equation}
We now assume that there is a finite number of domain walls in the system. The positions of the domain walls are given by zeros of spatially varying magnetic field $h \left({\bf r}\right)$, which we assume to be obeying the Gaussian distribution. The procedure of averaging over the magnetic field is given in the appendix to the paper. In the following, we outline it's steps. First, we consider a locally flat domain boundary. This can be justified if the scale along the wall $L_{\|}\sim\sqrt{\frac{K}{\epsilon_{0}}}\sim\frac{1}{p_{F}\sqrt{\epsilon_{0}}}$ is much smaller than the domain wall curvature. Second, we consider domain walls to be independent from each other, meaning that overlap of eigenfunctions (\ref{psi-0}) of neighboring domain walls is exponentially small. This allows to separately average over the domain wall direction and position, and to introduce concentration of the walls $n_{W}$ in the definition of the domain wall susceptibility (\ref{chi-omega}) as $\chi \rightarrow \frac{n_{W}}{\beta}\chi$. The averaging of considered quantities over random values of $\epsilon_{0}$ does not need a cutoff at small $\epsilon_{0}$ and might be approximated by substituting the average value of $\epsilon_{0}$ in to the susceptibility.

According to (\ref{chi-omega}), the relaxation time of domain wall fluctuations is proportional to $\epsilon_{0}^{-1}$ and is much larger than the relaxation time of the volume fluctuations, which is proportional to $\alpha^{-1}$. At small values of $\epsilon_{0}$ (see expression (\ref{lifetime})), contribution of the domain wall fluctuations to the total susceptibility of the system can be approximated as
\begin{equation}
\delta\chi\sim n_{W}\int dx dx'\Psi_{0}\left( x\right) \Psi_{0}\left( x^{\prime}\right)
\chi(0,0)\sim \frac{\chi_{0}n_{W}}{\beta\epsilon_{0}}.
\end{equation}
And depending on the parameters, can be of the same order as volume susceptibility $\frac{\chi_{0}}{\alpha}$ when the ratio of domain wall concentration to the domain wall thickness $\frac{n_{W}}{\beta}$ is of order of $\frac{\epsilon_{0}}{\alpha}$.

\section{Resistivity temperature dependence}

Let us consider contribution to the resistivity due to domain wall scattering of conducting electrons. We will consider three temperature dependent contributions to the resistivity. They are due to scattering of conducting electrons by fluctuations of domain wall, due to domain walls shape change with the temperature variation, and the third one is due to variation of concentration of domain walls $n_{W}$. First two contributions are considered in this section and have a common nature. Third contribution depends on position of the system on the phase diagram. We discuss its contribution in the conclusions of the paper.  Contribution to the resistivity due to a mechanism of electron scattering by spin fluctuation is obtained in second order perturbation theory in interaction (\ref{coupling}) and is expressed through imaginary part of averaged susceptibility as \cite{Ueda, Moriya}
\begin{equation}
\rho
(T)=R_{0}\frac{1}{T}\int\limits_{0}^{2p_{F}}\frac{dqq^{3}}{p_{F}^{4}}
\int\limits_{-\infty}^{\infty}\frac{d\omega\omega}{\sinh^{2}\left(
\omega/2T\right)}\operatorname{Im}\chi\left( {\bf q},\omega\right),
\end{equation}
here $R_{0}=\frac{m_{e}G^{2}\nu}{32e^{2}n}$, where $\nu$ is density of states of conducting electrons per spin, $p_{F}$, $n$ are Fermi momentum, density of conduction electrons respectively.

In addition, the fluctuations give temperature dependent contribution to the average magnetization of the domain wall. Fluctuation part of magnetization is determined by a derivative $\delta m(\textbf{r})=-\frac{\delta\Delta \Omega}{\delta h(\textbf{r})}$ of fluctuation part of the free energy
$\Delta\Omega=\frac{1}{2}T\sum_{\omega_{n},Q}\ln\left(\epsilon + \Gamma(Q,\omega_{n})\right)$. This leads to a change of the domain wall profile and gives additional temperature dependence of resistivity, which is proportional to $G^{2}m(\textbf{Q})\delta m(\textbf{Q},T)$. The sum of both contributions to the resistivity is then given by
\begin{eqnarray}
&R=\frac{4\pi \beta^2 R_{0}}{p_{F}^{4}}\int\limits_{0}^{\infty}
d\omega\left(\frac{\omega}{T\sinh^{2}(
\omega/2T)}-2\coth(\frac{\omega}{2T}) +2
\right)\nonumber\\
&\times \int\frac{d^{2}Q}{(2\pi)^{2}}\operatorname{Im}\chi(\textbf{Q},\omega).
\end{eqnarray}
Term, proportional to $(-\coth(\frac{\omega}{2T}) +1)$, is due to $\delta m(\textbf{Q},T)$. The domain wall susceptibility here is given by the expression (\ref{chi-omega}) with a substitution $\chi \rightarrow \frac{n_{W}}{\beta}\chi$ already made.

Our calculations show that despite of the different expression for damping in ballistic (\ref{ballistic damping}) and diffusive (\ref{diffusive damping}) regimes, the temperature dependence of resistivity has the same analytical form for both of them. It is quadratic at low temperatures and linear at larger. The transition temperature is proportional to the inverse relaxation time of the domain walls fluctuations.
For the ballistic scattering regime at temperatures $T<T_{0}$ the resistivity has a quadratic temperature dependence
\begin{equation}\label{quadrat}
R=\frac{2\pi \beta
R_{0}n_{W}\chi_{0}}{p_{F}^{4}}\frac{\gamma T^{2}}{KT_{0}},
\end{equation}
and at higher temperatures $T>T_{0}$, the dependence becomes linear
\begin{equation}\label{linear}
R=\frac{2\pi \beta R_{0}n_{W}\chi_{0}}{p_{F}^{4}}\frac{T}{K},
\end{equation}
here $T_{0}= \gamma \beta v_{F}\epsilon_{0}/(5\ln(\beta^{2}K/\epsilon_{0}))$. In the case of diffusive scattering we get that the resistivity temperature dependence has the same form with $T_{0}$ modified to $T_{0} =\gamma D\beta\epsilon_{0}\sqrt{K/\epsilon_{0}}$. The transition from quadratic $T^{2}$ to linear $T$ dependence corresponds to transition from quantum to thermal fluctuations of domain wall. \cite{Onuki}

We would like to notice that the same temperature dependence holds for scattering by volume spin
fluctuations, except for the difference of effective $T_{0}$ in ballistic and diffusive cases. \cite{Fulde, Jull} Let us compare obtained contribution to the resistivity (\ref{quadrat}) with one due to volume spin
fluctuations $\varrho_{vol}(T)$. In the considered temperature range, the contribution of volume spin-fluctuations is quadratic and is given by $\varrho_{vol}(T) \sim R_{0}\chi_{0}T^{2}/(p_{F}v_{F}\sqrt{\alpha})$. \cite{Ueda, Moriya} Therefore, ratio of contributions of domain walls fluctuations to volume spin fluctuations,  $R(T)/\varrho_{vol}(T) \sim \frac{n_{W}}{p_{F}\epsilon_{0}}$, can be of order of one.

\section{Interaction correction}

At low temperatures, an important resistivity temperature dependence is related to weak localization and electron interaction corrections.\cite{AlAr} Here we are going to discuss contribution to the conductivity originating from the interplay between electron inelastic scattering by domain wall fluctuations and elastic scattering by impurities. The triplet channel contribution to the conductivity after disorder averaging is given by \cite{Zala, Paul}
\begin{eqnarray}\label{alar}
&\delta\sigma = 2\pi e^2 v_{F}^2 \tau \nu G^2 \int
\frac{d\omega}{4\pi^2}\left[\frac{\partial}{\partial\omega}
\left( \omega\coth\frac{\omega}{2T} \right)\right] \nonumber \\
&\times \operatorname{Im}\int \frac{d^{3}q}{(2\pi)^{3}}B({\bf q}, \omega) \chi({\bf q}, \omega),
\end{eqnarray}
here we also averaged over the positions of the domain walls after which the equation above became isotropic. In the following, we will only be interested in the temperature dependent terms of the contribution (\ref{alar}).

In the ballistic regime $T \tau > 1$ one can use the following approximation $B({\bf q}, \omega) \approx 2/(v_{F}q)^2$, which is valid for $v_{F}q > |\omega|$. Expression for the susceptibility of the domain
wall (\ref{chi-omega}) at small momenta $q < \beta$ is $\chi({\bf q}, \omega) =
4n_{W}\chi_{0}/\beta\left(\epsilon_{0} + KQ^2 + i(4\gamma\omega/(\pi v_{F}\beta)) \ln\left(\alpha/KQ^2\right)
 \right)^{-1}$.

We find that at temperatures $T>\pi v_{F} \beta\epsilon_{0}/(2\gamma)\simeq T_{0}$ and $T<v_{F}\sqrt{\epsilon_{0}/K}$, the contribution to the conductivity is logarithmic in temperature
\begin{equation} \label{interactioncorrection1}
\delta\sigma = -\left(\frac{\chi_{0}}{2\pi^2\gamma} e^{2} \nu G^{2} n_{W} \right)\sqrt{\epsilon_{0}/K}v_{F}\tau \ln\left(\frac{\sqrt{\epsilon_{0}/K}v_{F}}{T}\right).
\end{equation}
At higher temperatures $T > v_{F}\sqrt{\epsilon_{0}/K}$ and $T < \frac{2 v_{F} \gamma}{\pi K \beta}$, the contribution to the conductivity becomes linear
\begin{equation}\label{interactioncorrection2}
\delta\sigma = \left(\frac{5\chi_{0}}{12\pi^2\gamma} e^{2} \nu G^{2} n_{W} \right) \tau T .
\end{equation}
As the temperatures increases further, the contribution decays as $1/T$.

Next, let us discuss the diffusive regime at $T \tau < 1$. In this case $B({\bf q}, \omega) = \frac{4}{3}\frac{Dq^2}{(Dq^2 + i\omega)^3}$ and $\chi({\bf q}, \omega) = 4n_{W}\chi_{0}/\beta\left(\epsilon_{0} +
KQ^2 + i2\gamma\omega /(D\beta Q) \right)^{-1}$ are approximated at small momenta $q<\beta$. Most singular contribution arises from $\sqrt{|\omega|/D}<q<\min(\beta,1/\ell)$, and $q>\sqrt{\epsilon_{0}/K}$. Where $\ell$ is
the electron`s mean free path. Calculations show that at temperatures $p=\min(\Omega_{1},\Omega_{2},1/\tau) > T$, where $\Omega_{1}=\frac{\gamma^{2}D}{2K^2\beta^2}$, $\Omega_{2} = \sqrt{2}\beta^{4}DK /\gamma$ the contribution to the conductivity is given by
\begin{equation}\label{interactioncorrection3}
\delta\sigma = -\left( \frac{2^{7/2}\chi_{0}}{9\pi^3 \gamma}e^{2}\nu G^{2} n_{W}\right) \ln\left(\frac{p}{T}\right)\ln\left(\frac{pT}{\Omega_{1}^{2}}
\right).
\end{equation}
Expressions (\ref{interactioncorrection1}), (\ref{interactioncorrection2}), and (\ref{interactioncorrection3}) have a two-dimensional like \cite{AlAr,Paul} temperature behavior while obtained for a three-dimensional interacting electron system. To mention, all of the expressions have a non-Fermi liquid contribution to the conductivity. One can relate the obtained results for the resistivity as $\delta\rho = (1 - \delta\sigma/\sigma_{D})/\sigma_{D}$, where $\sigma_{D}$ is the Drude conductivity. We would like to point out that we have not considered all of the possible regimes of parameters focusing on the most interesting ones.

\section{Conclusions}

To conclude, we have proposed possible scenario of a non-Fermi liquid temperature behavior of resistivity of a metallic metamagnet undergoing first order phase transition. The present paper is motivated by experiments (see introduction of the paper), where it was observed that a system close to a putative quantum critical point shows a non-Fermi liquid temperature dependence of resistivity in a wide range of parameters, such as temperature, pressure or magnetic field. In this case the theory \cite{Moriya, Hertz, Millis} of electron scattering on critical spin-fluctuations fails to explain the experimental observations. In the present paper we point out on the importance of the phase separation in the properties of a system close to putative quantum critical point. In studied model, the metamagnet undergoes a phase separation to magnetic domains with corresponding domain walls.    It is shown that in the case of weak pinning, when $\alpha >> \epsilon_{0}$ (see expression (\ref{lifetime})), the relaxation time of domain wall fluctuations is much larger than the relaxation of volume spin density. It is important that in this regime the domain wall fluctuations are more critical than the volume fluctuations. Therefore, the scattering of conducting electrons on the these domain wall fluctuations results in non-Fermi liquid dependence of resistivity, which is linear in temperature (see expression (\ref{linear})). In this temperature range, the contribution of electron scattering on volume fluctuations is quadratic in temperature (of a Fermi liquid type).\cite{Ueda, Moriya} And as we have shown, the non-Fermi liquid contribution (\ref{linear}) can be a dominant one when the density of domain walls increases. We also considered the interaction correction to the conductivity due to an interplay of impurity and spin fluctuations scattering. Overall, expressions (\ref{linear}), (\ref{interactioncorrection1}), (\ref{interactioncorrection2}), and (\ref{interactioncorrection3}) are main results of the presented paper.

Let us discuss how contribution related to the dynamics of domain walls might be the dominating one.
In the temperature range considered in this paper, the contribution of volume fluctuations to the resistivity is always $\sim T^{2}$. It is reasonable to assume that additional resistivity due to elastic scattering of conduction electrons by domain walls is proportional to $n_{w}$. Depending on position of average magnetic field relatively to a metamagnetic value, $h_{m}(T)$ (see section \Rmnum{2} for definition), the concentration $n_{w}$ can increase or decrease with the temperature. For example, when $h_{m}(T)$ is an increasing function of temperature, the $n_{W}$ decreases with temperature if average magnetic field is smaller than $h_{m}(T=0)$. In considered temperature range, when dependence $h_{m}(T)$ on temperature is determined by volume fluctuations, the concentration $n_{w}$ quadratically changes with the temperature. In case when $n_{w}$ decreases with temperature there can be a cancelation of quadratic temperature
dependence of volume contribution to the resistivity by the contribution of domain walls $n_{W}(T)$.
In case of the cancelation, obtained results (\ref{linear}), (\ref{interactioncorrection1}), and (\ref{interactioncorrection2}) will be dominant. And the total resistivity will have quasi-linear non-Fermi liquid temperature dependence.

One of the assumptions we made in this paper is the flatness of the domain walls. Another important point is related to the type of the magnetic domain structure. We considered magnetic domain walls, which are independent of each other with random positions described by a Gaussian distribution. Possible arrangement of domain walls in to periodic structure, such as striped or hexagonal, will certainly change the dynamical susceptibility of the spin-fluctuations, and therefore contribute differently to temperature behavior of the resistivity. We address the elaboration of these assumptions to future research.

\begin{acknowledgements}
This work was financially supported by ARO grant W911NF-09-1-0527, NSF grant DMR-0955778, and RFFI grant
12-02-00300-A.
\end{acknowledgements}

\appendix
\section{Averaging over random surfaces}

Here we discuss the procedure of averaging over the random positions of domain walls and random values of $\epsilon_{0}$. We assume that magnetic field is described by a Gaussian distribution, such that
\begin{equation}\label{corr}
\langle \Delta h(\textbf{r})\Delta h(\textbf{r}^{\prime}) \rangle
=A\exp(-\zeta (\textbf{r}-\textbf{r}^{\prime})^{2})
\end{equation}
and $\langle \Delta h(\textbf{r}) \rangle=0$, where $\Delta h(\textbf{r})=h(\textbf{r})-h_{m}$. When parameter $\zeta$ is small, the magnetic field becomes a slowly varying random function $ h(\textbf{r})$. Typical phase boundary in such a random field is smooth. Positions of the domain walls are defined by the following equation
\begin{equation}\label{sur-def}
h(\textbf{r})=h_{m}+\Delta h(\textbf{r})=0.
\end{equation}
Note, that under this definition we neglect cases when in small closed regions there is no solution of mean field equation (\ref{mean-field}) for favorite phase, or energy associated with it is too high.

Fourier transform of any quantity $V(\textbf{r})$ which is nonzero near the domain wall surface and slowly varying along it, is calculated as
\begin{widetext}
\begin{equation}\label{sur-int}
V(\textbf{q})=\int d\textbf{r}~ \exp(i\textbf{qr})
V(\textbf{r})
=\int dS~ \exp(i\textbf{q}\textbf{r}_{S}) V(\textbf{q}
\textbf{n} (\textbf{r}_{S}),\textbf{r}_{S})
=\int d\textbf{r}~
V(\textbf{q}\textbf{n}(\textbf{r}),\textbf{r})\exp(i\textbf{q}\textbf{r})|\frac{d}{d\textbf{r}}
h(\textbf{r})|\delta(h(\textbf{r})),
\end{equation}
here $\textbf{r}_{S}$ is a point on the surface, and $\textbf{n}(\textbf{r}_{S})$ is a vector normal to the surface at the point $\textbf{r}_{S}$ defined as $\textbf{n}(\textbf{r})=\frac{dh(\textbf{r})}{d\textbf{r}}/|\frac{d}{d\textbf{r}} h(\textbf{r})|$.
In (\ref{sur-int}) $V(\textbf{q} \textbf{n}(\textbf{r}),\textbf{r})$ is a one dimensional Fourier transform in the direction normal to the surface.

We can now average the dynamical susceptibility of the number of domain wall. For one domain wall the susceptibility is given by the expression (\ref{dyn-sus}). The scale along the surface of the domain wall that we are interested in is $L_{\|}\sim\sqrt{\frac{K}{\epsilon_{0}}}\sim\frac{1}{p_{F}\sqrt{\epsilon_{0}}}$, here $\delta h\sim\sqrt{A}\sim\frac{\alpha m_{0}}{\chi_{0}}$ and
the $\epsilon_{0}$ is estimated as
\begin{equation}
\epsilon_{0}\sim \frac{\chi_{0}}{m_{0}\sqrt{\alpha /K}}|\frac{dh}{dx}| \sim \sqrt{\alpha K}\sqrt{\zeta}.
\end{equation}
Therefore $L_{\|}\sim\frac{1}{p_{F}}\sqrt{\beta/\alpha\sqrt{\zeta}}$ is smaller than the radius of a surface curvature $L_{\|}\sqrt{\zeta}\sim
\sqrt{\sqrt{\zeta}/\beta}<<1$, and under averaging procedure we can use expression (\ref{dyn-sus}). Then the average susceptibility is
\begin{equation}\label{dynamic1}
\chi(\textbf{q},\omega) =\int d^3\textbf{R}\int
d\textbf{b} \Pi(\textbf{R}, \textbf{b})
\exp(i\textbf{q}\textbf{R}) \textbf{b}^{2}
\Psi_{0}(\textbf{q}\textbf{n})\Psi_{0}(-\textbf{q}\textbf{n})\int \frac{d^2
Q}{(2\pi)^2} \exp(i\textbf{QR})\chi(\textbf{Q},\omega),
\end{equation}
where $\Psi_{0}(q)=2\beta^{3/2}/(q^2+\beta^{2})$ is a Fourier transform of $\Psi_{0}(x)$ defined by (\ref{psi-0}), and
\begin{equation}
\Pi(\textbf{r}_{1}-\textbf{r}_{2}, \textbf{b})=\langle \delta(
h(\textbf{r}_{1}))\delta( h(\textbf{r}_{2}))
\delta(\textbf{b}-\frac{1}{2}(\frac{d}{d\textbf{r}_{1}}
h(\textbf{r}_{1})+\frac{d}{d\textbf{r}_{2}} h(\textbf{r}_{2})))
\rangle .
\end{equation}
\end{widetext}
At $R << \sqrt{\zeta}$ the $\textbf{b}$ is a normal to the surface and therefore is also normal to $\textbf{R}$ . Therefore in
(\ref{dynamic1}) we can separately average over the direction and the value of $\textbf{b}$, considering $\Pi(\textbf{R}, \textbf{b})$ only in
the limit of $R\sqrt{\zeta}<<1$. In case of the Gaussian distribution, the $\Pi(\textbf{R}, \textbf{b})$ is calculated analytically as
\begin{equation}
\Pi(\textbf{R},\textbf{b})=\frac{n_{W}}{16RA\zeta}P(\textbf{b}_{\bot})P(\textbf{b}_{\|}),
\end{equation}
where
\begin{equation}
P(\textbf{b}_{\bot})=\frac{1}{2\pi
A\zeta}\exp(-\frac{\textbf{b}_{\bot}^{2}}{2A\zeta})
\end{equation}
and
\begin{equation}
P(\textbf{b}_{\|})=\sqrt{\frac{3}{\pi A \zeta ^3 R^4}}
\exp(-\frac{3\textbf{b}_{\|}^{2}}{A\zeta ^{3} R^{4}}),
\end{equation}
here $\textbf{b}_{\bot}$ and $\textbf{b}_{\|}$ are perpendicular and parallel components to $\textbf{R}$ consequently. The quantity $n_{W}$ has a meaning of the domain wall concentration and is defined as
\begin{equation}
n_{W}\equiv\frac{\langle \int d\textbf{r} \delta(
h(\textbf{r}))|\frac{d}{d\textbf{r}} h(\textbf{r})|
\rangle}{V}=\frac{\sqrt{8\zeta}}{\pi}\exp(-\frac{h_{m}^{2}}{A})
\end{equation}
where $V$ is the volume of the system. The factor $\textbf{b}^{2}$ under the integral of expression (\ref{dynamic1}) is estimated as $\textbf{b}^{2}\sim
\epsilon_{0}^2$. So averaging of quantities like susceptibility does not diverge at small $\epsilon_{0}$ and can be approximated by inserting average $\epsilon_{0}$.
Finally, the averaged susceptibility over the domain wall positions is given by the next expression:
\begin{equation}
\chi({\bf q},\omega) = \frac{1}{4\pi}\int d^2{\bf n} \Psi_{0}({\bf qn})\Psi_{0}({-\bf qn})\frac{n_{W}\chi_{0}}{\epsilon_{0} + K{\bf Q}^2 + \Gamma({\bf Q},i\omega)},
\end{equation}
where $\epsilon_{0}$ is approximated by its average value. At small momenta $q < \beta$ Fourier transform $\Psi_{0}({\bf qn})$ is approximated as $\Psi_{0}=2/\sqrt{\beta}$.

\end{document}